\documentclass[onecolumn]{aastex61}   

\newcommand{\about}{\mbox{$\sim$}}               
\newcommand{\Lbol}{\mbox{$L_{\rm bol}$}}                    
\newcommand{\Lir}{\mbox{$L_{\rm 8-1000\; \mu m}$}}  
\newcommand{\Lsol}{\mbox{$L_\odot$}}             
\newcommand{\Lsun}{\mbox{$L_\odot$}}            
\newcommand{\Msol}{\mbox{$M_\odot$}}            
\newcommand{\Msun}{\mbox{$M_\odot$}}           
\newcommand{\NH}{\mbox{$N_{\rm H}$}}             
\newcommand{\Tb}{\mbox{$T_{\rm b}$}}               
\newcommand{\Tdust}{\mbox{$T_{\rm dust}$}}     
\newcommand{\uv}{\mbox{$u$--$v$}}                     
\newcommand{\hr}{\mbox{$^{\rm h}$}}                   
\newcommand{\mn}{\mbox{$^{\rm m}$}}                
\newcommand{\perbeam}{\mbox{beam$^{-1}$}}                         
\newcommand{\persquarecm}{\mbox{cm$^{-2}$}}                      
\newcommand{\kms}{\mbox{km s$^{-1}$}}                                    
\newcommand{\persquarekpc}{\mbox{kpc$^{-2}$}}                    
\newcommand{\persquarepc}{\mbox{pc$^{-2}$}}                         

\newcommand{\micro}{\mbox{$\mu$}}    
\newcommand{\twelveCO}{\mbox{$^{12}$CO}}                  
\newcommand{\thirteenCO}{\mbox{$^{13}$CO}}                
\newcommand{\CeighteenO}{\mbox{C$^{18}$O}}              
\newcommand{\HCthreeN}{\mbox{HC$_{3}$N}}                 
\newcommand{\HCfiveN}{\mbox{HC$_{5}$N}}                    
\newcommand{\propyne}{\mbox{CH$_3$CCH}}                 
\newcommand{\CHthreeCCH}{\mbox{\propyne}}                 
\newcommand{\CHthreeCN}{\mbox{CH$_3$CN}}                 
\newcommand{\CthirtyfourS}{\mbox{C$^{34}$S}}              

\newcommand{\nd}{\nodata}
\newcommand{\tnm}[1]{\tablenotemark{#1}}
\newcommand{\citest}[1]{\citeauthor*{#1}}
\newcommand{\citesp}[1]{(\citeauthor*{#1})}

\newcommand{\CDCone}{\mbox{CDC$_1$}}      
\newcommand{\CDCtwo}{\mbox{CDC$_2$}}      
\newcommand{\Eone}{\mbox{E$_1$}}           
\newcommand{\Etwo}{\mbox{E$_2$}}           
\newcommand{\Wone}{\mbox{W$_1$}}          
\newcommand{\Wtwo}{\mbox{W$_2$}}          
\newcommand{\Tp}{\mbox{$T_{\rm p}$}}      
\newcommand{\rmaj}{\mbox{$r_{\rm maj}$}} 
\newcommand{\alphad}{\mbox{$\alpha_{\rm d}$}} 
\newcommand{\alphap}{\mbox{$\alpha_{\rm p}$}} 
\newcommand{\fdust}{\mbox{$f_{\rm d}$}}    
\newcommand{\Eoneprime}{\mbox{E$_1$$'$}}           
\newcommand{\Woneprime}{\mbox{W$_1$$'$}}         
\newcommand{\Etwoprime}{\mbox{E$_2$$'$}}           
\newcommand{\Wtwoprime}{\mbox{W$_2$$'$}}         

\shorttitle{Structure of Arp 220 Nuclei at 3 mm}
\shortauthors{Sakamoto et al.}

\accepted{for publication in ApJ}

\begin{document}
\title{Resolved Structure of Arp 220 Nuclei at $\lambda \approx 3 $ mm}

\author{Kazushi Sakamoto}
\affiliation{Academia Sinica, Institute of Astronomy and Astrophysics, Taipei 10617, Taiwan}
 
\author{Susanne Aalto}
\affiliation{Department of Earth and Space Sciences, Chalmers University of Technology, Onsala Observatory, 439 92 Onsala, Sweden}

\author{Loreto Barcos-Mu\~{n}oz}
\affiliation{Joint ALMA Observatory, Alonso de C\'{o}rdova 3107, Vitacura, Santiago, Chile}
\affiliation{National Radio Astronomy Observatory, 520 Edgemont Road, Charlottesville, VA 22903, USA}

\author{Francesco Costagliola}
\affiliation{Department of Earth and Space Sciences, Chalmers University of Technology, Onsala Observatory, 439 92 Onsala, Sweden}

\author{Aaron S. Evans}
\affiliation{Department of Astronomy, University of Virginia, P.O. Box 400325, Charlottesville, VA 22904, USA}
\affiliation{National Radio Astronomy Observatory, 520 Edgemont Road, Charlottesville, VA 22903, USA}

\author{Nanase Harada}
\affiliation{Academia Sinica, Institute of Astronomy and Astrophysics, Taipei 10617, Taiwan}

\author{Sergio Mart\'{i}n}
\affiliation{Joint ALMA Observatory, Alonso de C\'{o}rdova 3107, Vitacura, Santiago, Chile}
\affiliation{European Southern Observatory, Alonso de C\'{o}rdova 3107, Vitacura, Santiago, Chile}

\author{Martina Wiedner}
\affiliation{LERMA, Observatoire de Paris, PSL Research University, CNRS, Sorbonne Universités, UPMC Univ. Paris 06, F-75014 Paris, France}

\author{David Wilner}
\affiliation{Harvard-Smithsonian Center for Astrophysics, 60 Garden Street, Cambridge, MA 02138, USA}

\correspondingauthor{K. Sakamoto}

\begin{abstract}
We analyze 3 mm emission of the ultraluminous infrared galaxy Arp 220 for spatially-resolved structure and spectral properties of the merger nuclei.
ALMA archival data at \about0\farcs05 resolution are used for extensive visibility fitting and deep imaging of continuum emission.
%
The data are fitted well with two concentric components for each nucleus, such as two  Gaussians 
or one Gaussian plus one exponential disk.
The larger components in individual nuclei are similar in shape and extent, \about100--150 pc, to the cm-wave emission due to supernovae. 
They are therefore identified with the known starburst nuclear disks. 
The smaller components in both nuclei have about a few 10 pc sizes and 
peak brightness temperatures (\Tb) more than twice higher than in previous single-Gaussian fitting.
They correspond to the dust emission that we find centrally concentrated in both nuclei 
by subtracting the plasma emission measured at 33 GHz. 
The dust emission in the western nucleus is found to have a peak $\Tb \approx 530$ K
and a full width at half maximum of about 20 pc.
This component is estimated to have a bolometric luminosity on the order of $10^{12.5}$ \Lsol\ and a 20 pc-scale 
luminosity surface density $10^{15.5}$ \Lsol\persquarekpc.
A luminous AGN is a plausible energy source for these high values while other explanations remain to be explored.
Our continuum image also reveals a third structural component of the western nucleus --- a pair of faint spurs 
perpendicular to the disk major axis.
We attribute it to a bipolar outflow from the highly inclined ($i \approx 60\degr$) western nuclear disk.
\end{abstract}  


\section{Introduction}   \label{s.introduction}
The nearest ultraluminous infrared galaxy \object{Arp 220} has been a key object in our study of
the luminous phase in galaxy evolution after a major merger \citep{Sanders96,Hopkins08}. 
It has two merger nuclei separated by about 1\arcsec (\about400 pc) on the sky \citep{Scoville98,Genzel01}
each having a \about100 pc scale rotating disks of molecular gas \citep{Sakamoto99}.
Vigorous star formation is evident in the nuclei from radio emission due to supernovae \citep[hereafter BM15]{Smith98,Barcos-Munoz15}.
Molecular outflows from the individual nuclei have been found \citep{Sakamoto09} as is often the case for luminous galactic nuclei 
\citep[e.g.,][]{Cicone14,Sakamoto14}.
Arp 220 is therefore undergoing merger-driven rapid evolution of galaxy nuclei.
Important open issues about the galaxy include the gas flows to, from, and within the two merger nuclei 
and the structure, physical and chemical properties, and the dominant luminosity sources of the nuclei.
On the last point, while nuclear starburst is evident whether there are any active galactic nuclei (AGN) with significant luminosities 
is still under intense study \citep[and references therein]{Yoast-Hull17,Paggi17} mainly because the nuclei are extremely obscured \citep[$\NH \geq 10^{25-26}$ \persquarecm,][hereafter S17]{Sakamoto08,Wilson14,Martin16,Scoville17}.
High-resolution observations at centimeter to sub-millimeter wavelengths are especially useful for many of the open issues thanks 
to the lower opacity compared to shorter wavelengths.

We recently found from our 1 and 0.8 mm observations of Arp 220 with the Atacama Large Millimeter-submillimeter Array (ALMA) 
that individual merger nuclei have composite structure (Sakamoto et al. in prep.). 
Each nucleus consists of a central compact core and a more extended structure that can be together fitted with two Gaussians. 
The single Gaussian models that had been used before are not adequate anymore for high quality data
at $\lesssim 0\farcs2$ resolution.
The central core components have sizes as small as 0\farcs1--0\farcs05 in full width at half maximum (FWHM)
and have higher peak brightness temperatures than estimated with single-Gaussian fits.
They are of great interest for unveiling the unknown nature of the luminosity source and to trace
the evolution of the merger nuclei. 
We have therefore extended our structural analysis of the Arp 220 nuclei to ALMA archival data obtained at around 3 mm  at \about0\farcs05 resolution. 
We have also decomposed the 3 mm continuum to plasma and dust emission and mapped the dust emission in individual nuclei.
This paper reports the results.

We will refer to the eastern nucleus of Arp 220 as Arp 220 E (sometimes just `E' for short) and the western nucleus
as Arp 220 W (or just W). 
All features within about 0\farcs5 of the centroid of each nucleus will be referred to in this way.
We adopt an angular size distance of $D_{\rm A} = 85.0$ Mpc (1\arcsec\ = 412 pc), luminosity distance $D_{\rm L} = 87.9$ Mpc,
and the total IR luminosity of \Lir = $10^{12.28} \Lsun$ for the galaxy \citep{Armus09} 
to be consistent with \citest{Scoville17}.

\section{ALMA Data} \label{s.almadata}

We analyzed archival data of the ALMA project 2015.1.00113.S (P.I.~Scoville). 
\citest{Scoville17} already reported a part of the project regarding \twelveCO(1--0) and continuum imaging
as well as spatial and spectral modeling of the CO emission.
The dataset consists of two tunings around 100 GHz consecutively observed on 2015 Oct.~27 
in a long-baseline configuration; \citest{Scoville17} analyzed the one with CO.
Both used J1550$+$0527 for flux and bandpass calibration. 
We adopted for it a flux model $(S_\nu / {\rm Jy}) = 1.000 \times (\nu/{\rm 104.0\; GHz})^{-0.602}$ that 
 is from the nearest records in the ALMA Calibrator Source Catalogue, i.e.,
 four measurements made at 91.5 and 233 GHz on Oct.~31 and Nov.~1.\footnote{
 The ALMA Observatory used the Morita Array (7m array) for the quasar measurements,
 tied them to the measurements of Neptune, Mars, and Titan on the same days, 
 and used the Butler-JPL-Horizons 2012 model for the solar system objects.
 The expected flux calibration accuracies in the ALMA Cycle 3 Proposer's Guide are (better than)  5\% and 10\% for 92 GHz and 233 GHz, respectively. 
 For our calibrator, two measurements at each frequency agree quite well. 
 As a result the formal 1$\sigma$ uncertainties of our calibrator model are 0.004 Jy for the flux density and 0.10 for the spectral index 
 after normalizing the power-law fit uncertainties so that the reduced $\chi^2$ is unity. 
 These uncertainties do not include any error due to deviation of the quasar spectral energy distribution from a power-law
 and model uncertainties of the solar system objects.
 The Observatory used for the archived data products 
 $(S_\nu / {\rm Jy}) = 1.126 \times (\nu/{\rm 104.0\; GHz})^{-0.680}$ for J1550+0527.
}
We flagged three poorly performing antennas, DA45, DA65, and DV20, and used the remaining 37.
The projected baseline lengths for Arp 220 ranged 0.24--10.38 km for the first tuning with \twelveCO\ 
and 0.22--11.10 km for the second with \thirteenCO.
We used CASA 4.7.2 for our data reduction \citep{CASA07},
starting from the raw data and following the steps of the observatory-provided calibration except the revised calibrator flux model
and the additional flagging of DA65 and DV20. 
We made phase-only self calibration using continuum.
It was first done independently for the two tunings, and we used that data for our visibility fitting.
Another round was made after combining the two datasets and before making the images presented in this paper.

\clearpage  
\section{3 mm Spectra of the Two Nuclei} \label{s.spectra}

\begin{figure*}[bht!]
\epsscale{1.18}
\plotone{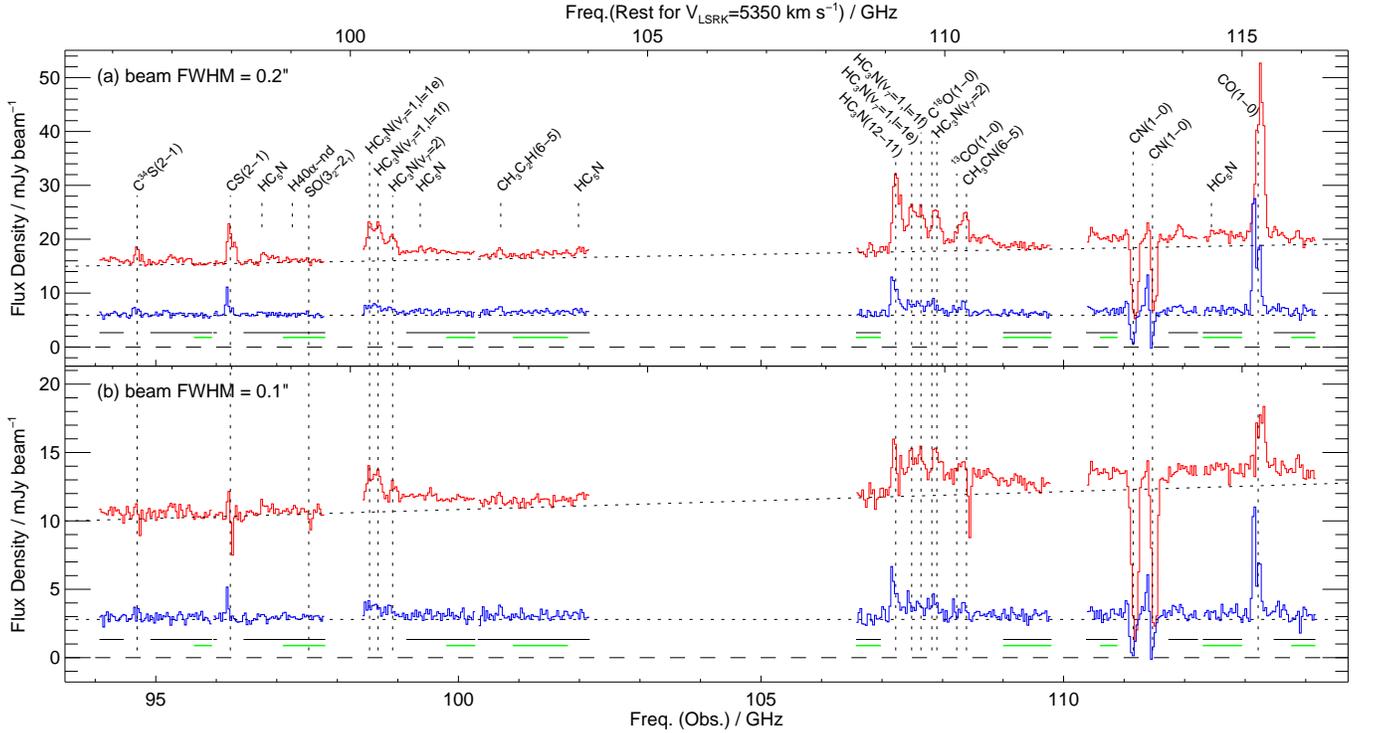}  
\caption{ \label{f.B3spec}
Spectra of Arp 220 nuclei around 100 GHz sampled at the western (red) and eastern (blue) nuclei.
Data imaged at every 10 MHz were binned to 30 MHz and convolved from about 0\farcs09 resolution to (a) 0\farcs2 and (b) 0\farcs1. 
Major lines are labeled with long vertical dotted lines.
Some minor lines are also labeled with short vertical lines.
Estimated power-law continuum is plotted for each nucleus as a black dotted line. 
Its power-law (i.e., spectral) index is $+1.2$ and $0.0$ for the western and eastern nucleus, respectively, in both panels.
Horizontal black and green bars below the spectra indicate the spectral segments that we analyzed as 
continuum-dominated channels. The black ones are used for continuum imaging and green for visibility fitting.
}
\end{figure*}

Figure~\ref{f.B3spec} shows the spectra sampled at individual nuclei with 0\farcs2 and 0\farcs1 beams.
Major lines in the spectra are \twelveCO(1--0), CN(1--0) doublet, \CHthreeCN(6--5),
\HCthreeN(12--11), \HCthreeN(11--10), CS(2--1), and \CthirtyfourS(2--1).
The \HCthreeN\ lines include transitions within vibrationally excited states $v_7=1$ ($l=1e$ and $1f$) and $v_7=2$.
\thirteenCO(1--0) and \CeighteenO(1--0) are also in the frequency coverage but they are blended 
with other lines and inconspicuous.
Other likely identifications are \CHthreeCCH\ and \HCfiveN\ while H40$\alpha$ is not detected.
These lines and more have been detected in emission in single-dish observations by \citet{Aladro15}.
In the ALMA high-resolution data here, the CN doublet is almost totally in absorption 
while \CHthreeCN(6--5), CS(2--1), \CthirtyfourS(2--1), and SO(3$_2$--2$_1$)
start to show absorption toward Arp 220 W at 0\farcs1 resolution. 
While we will measure continuum source sizes later, this observation 
already indicates that the western nucleus has a bright continuum source or sources whose extent  $d$ satisfies 
$d \lesssim 0\farcs1$ and $d \ll\hspace{-1.0em}/\ 0\farcs1$,
because lines would have been in absorption at both resolutions if $d \gtrsim 0\farcs2$
and in emission if $d \ll 0\farcs1$.

%
Continuum was first estimated by visually finding regions with least lines in each spectrum and fitting a power-law curve 
$S_\nu \propto \nu^\alpha$;
off-line channels around 96--98, 107, and 109.5 GHz constrained the fit most.
Obtained curves are plotted in Fig.~\ref{f.B3spec} and have a power-law (i.e., spectral) indices 
$\alpha \equiv d \log S_\nu/d \log \nu$ of
$+1.2$ for the western nucleus and $0.0$ for the eastern nucleus with uncertainties about $\pm 0.2$.
It is evident that the two nuclei have different spectral indices around 100 GHz.
Arp 220 W has a positive index indicating significant contribution from thermal dust emission or
partially opaque free-free emission. 
Arp 220 E has a nearly flat spectrum indicating much less contribution of such emission. 
Guided by these continuum fits and the observed spectra, we defined two sets of continuum-dominated channels (CDCs) 
for further analysis.
The first (\CDCone) is marked with black horizontal bars in Fig.~\ref{f.B3spec}. 
They are more than 650 \kms\ from the systemic velocity for major lines 
but include week possible lines such as \HCfiveN. The total bandwidth of \CDCone\ is 9.5 GHz.
Contribution of line emission to the integrated emission of \CDCone\ is in the range of 5--10\% for the two apertures and two nuclei. 
The second set of channels (\CDCtwo) is  marked with green bars in Fig.~\ref{f.B3spec}.
It is a subset of \CDCone\ and avoids some weak lines.

\vspace{6mm} 
\section{Visibility Fitting} \label{s.visfit}
	We made model fitting of the calibrated visibilities to obtain continuum parameters of the two nuclei. 
Visibility fitting is a powerful way to analyze interferometric data when the target structure is simple and marginally resolved
\citep[e.g.,][]{Wiedner02,  Sakamoto13}.\footnote{
 For example, a Gaussian source with 50 mas FWHM on the sky has a Gaussian-shape distribution of visibility amplitude 
 centered at the origin of the uv plane, with the amplitude declining to 50\% and 6\% of the central value (= total flux) 
 at the uv radii of 0.9 and 1.8 M$\lambda$, respectively. 
 Our data coverage to about 3 M$\lambda$ is sufficient to determine the amplitude distribution in the uv domain 
 and hence the source size in the image domain, provided that visibilities have sufficient signal-to-noise ratio.
 While our dataset has a beam size of about 50 mas with the uniform weighting of visibilities, even a smaller size can be measured from 
 a precise amplitude-to-uv radius curve \citep{Marti-Vidal12}.
 }
For such cases, it is far more straightforward to fit visibilities than to make a dirty image first, deconvolve it next by iteratively finding clean components and convolving them with a clean beam, 
and then make image-domain deconvolution of the cleaned image. 
We made our non-linear visibility fitting in IDL using an implementation of the Levenberg-Marquardt algorithm called {\tt mpfit} 
\citep{More77,More93,Markwardt09}.

\begin{deluxetable*}{lCCCCl}[h]
\tablecolumns{4}
\tablewidth{0pt}
\tablecaption{Arp 220 Parameters from 3 mm Continuum \label{t.ParamSummary} }
\tablehead{ 
       \colhead{Parameter} & 
       \multicolumn{2}{c}{Arp 220 E} &
       \multicolumn{2}{c}{Arp 220 W} &
       \colhead{unit}  
       }
\startdata
\sidehead{1 Gaussian fit}
major axis FWHM                                         & 227 \pm 5             & \nd & 83.5 \pm 1.2         & \nd & mas \\
minor axis FWHM                                         & 111 \pm 1             & \nd & 64.6 \pm 1.1         & \nd & mas \\
axial ratio (min./maj.)                                 & 0.453 \pm 0.007 & \nd & 0.765 \pm 0.010 & \nd & \\
major axis P.A.                                            & 47.4 \pm 1.0        & \nd & 110.4 \pm 1.1      & \nd & \degr \\
peak \Tb                                                      & 37.3 \pm 1.4        & \nd & 385.5 \pm 7.4      & \nd & K \\
R.A.(ICRS)\tnm{a} 15\hr34\mn               & 57.2917               & \nd & 57.2224               &\nd & sec  \\
Dec.(ICRS)\tnm{a} +23\degr30\arcmin & 11.337                  & \nd & 11.500                 &\nd &  \arcsec   \\
\sidehead{2 Gaussian fit}
component name                             & \Eone                   & \Etwo                    & \Wone                     & \Wtwo                     &  \\
major axis FWHM                              & 87.9\pm3.6        & 369 \pm 9             & 50.0 \pm0.7         & 232 \pm 3            & mas \\ 
minor axis FWHM                              & 43.5\pm3.5        & 188 \pm 5             & 36.0 \pm0.5         & 144 \pm 2             & mas \\
axial ratio (min./maj.)                      & 0.471\pm0.033 &  0.505 \pm 0.011 & 0.717 \pm 0.010 & 0.617 \pm 0.012 &  \\
major axis P.A.                                  & 55.0 \pm 2.4      & 49.8 \pm 1.1         & 136.1 \pm 1.9      & 83.3 \pm 0.4        & \degr \\
peak \Tb                                            & 72.8 \pm 4.6      & 14.7 \pm 0.7         & 639.9 \pm 13.7   & 45.0 \pm 1.5        & K \\
R.A. offset\tablenotemark{b}         & -9.6 \pm 0.8       & 12.0 \pm 1.1         &  1.3 \pm 0.1          & -11.1 \pm 0.7      & mas \\
Dec. offset\tablenotemark{b}        & -1.4 \pm 1.2        & -1.2 \pm 1.5         & -1.2 \pm 0.2         & 1.6 \pm 0.6           & mas\\
spectral index $\alpha$  \tnm{c}   &  -0.96 \pm 0.92   & 0.81 \pm 0.29  & 1.94 \pm 0.16     & 1.01 \pm 0.25   & \\
\sidehead{1Gaussian + 1Exp-disk fit\tablenotemark{d}}
component name                         & \Eone$'$                  & \Etwo$'$             & \Wone$'$            & \Wtwo$'$                    &  \\
major axis size \tablenotemark{e}  & $\lesssim 70$     & {\it 82.1}            & 55.7 \pm1.3       & {\it 56.9}               & mas \\ 
minor axis size \tablenotemark{e}  & $\lesssim 50$     & {\it 43.6}            & 32.9 \pm1.6       & {\it 33.9}               & mas \\
axial ratio (min./maj.)                & \nd                        &  {\it 0.531}         & 0.583 \pm 0.020     & {\it 0.596}            &  \\
major axis P.A.                            & \nd                        & {\it 54.7}            & 144.8 \pm 1.9          & {\it 79.4}               & \degr \\
peak \Tb                                      & \nd                       &  \nd                       & 529 \pm 20              & \nd                          & K \\
R.A. offset\tablenotemark{b}  & \nd                        & {\it 0}                   & 0.67 \pm 0.16         & {\it 0}                     & mas \\
Dec. offset\tablenotemark{b} & \nd                        & {\it 0}                   &  -0.97 \pm 0.09       & {\it 0}                     & mas\\
spectral index $\alpha$            &  \nd                      & {\it -0.59}             & 3.57 \pm 0.21\tnm{f}  & {\it -0.61}              & \\
\enddata
\tablecomments{
Parameters were obtained with visibility fitting.
Fitting results from the nine segments of continuum-dominated channels (\CDCtwo) were averaged using the inverse square
of their uncertainties as weights.
Uncertainties of the means here are $\pm 1\sigma$ and do not include any systematic errors. 
The peak (Rayleigh-Jeans) brightness temperatures are subject to the flux calibration uncertainty on the order of 5\% in this ALMA band.
}
\tablenotetext{a}{Absolute astrometry is estimated to be accurate to 5 mas from the
visibility fit positions of a test source J1532$+$2344 in the same observations.
}
\tablenotetext{b}{Offset from the 1-Gaussian fit position.}
\tablenotetext{c}{Unaccounted errors due to low-level line contamination are expected for components with low
brightness temperatures.}
\tablenotetext{d}{Parameters for the exponential disks \Etwo$'$ and \Wtwo$'$, in italics, are from \citest{Barcos-Munoz15} and fixed. }
\tablenotetext{e}{Gaussian FWHM for \Eoneprime\ and \Woneprime\ and exponential scale length for \Etwo$'$ and \Wtwo$'$.
For comparison between a Gaussian and an exponential disk, 
the half-light diameter of a Gaussian is its FWHM and that of an exponential disk is about 3.36 times the exponential scale length.}
\tablenotetext{f}{This spectral index is a consequence of an assumption about the spectral index of dust emission (\S \ref{s.f_dust}). 
See \S \ref{s.dustEmission} for a caution about the assumption.}
\end{deluxetable*}


\subsection{1G fit} \label{s.visfit.1G}
We first fitted the visibilities using one Gaussian component for each nucleus; we call this the 1G fit.
The whole data set was fitted in each and every 30 MHz channel.
The resulting parameters, averaged in \CDCtwo\ to minimize line contamination, are listed in Table \ref{t.ParamSummary}.
Our 1G-fit positions agree very well with those measured at 33 GHz by \citest{Barcos-Munoz15}.
Our 1G-fit parameters are consistent with the deconvolved parameters of \citest{Scoville17} for Arp 220 E
but we obtained a smaller size ($0\farcs084 \times 0\farcs065$)  and higher peak brightness temperature (386 K) 
for the more compact Arp 220 W,
for which \citest{Scoville17} obtained $0\farcs12 \times 0\farcs11$ for the deconvolved FWHM and 167 K for the peak deconvolved \Tb.
Our visibility-fit size agrees with our image-domain deconvolved size from a 0\farcs07 resolution image made with  {\tt robust=0.5}.

\subsection{r2G fit} \label{s.visfit.r2G}
\begin{figure*}[h]
\epsscale{1.17}
\plottwo{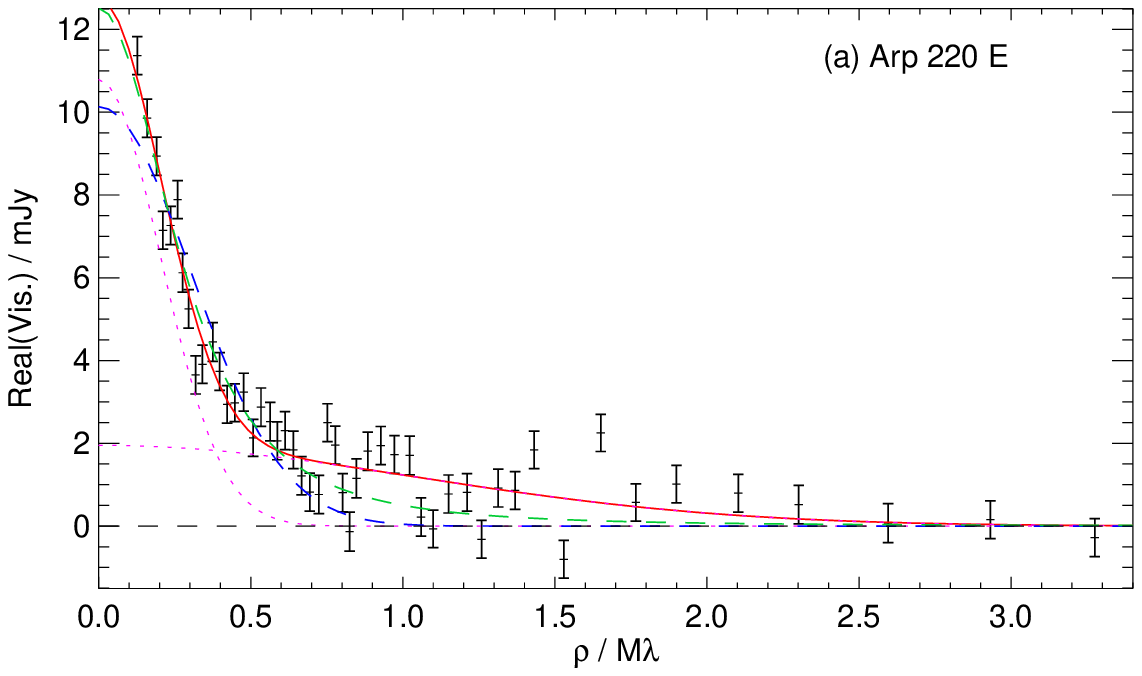}{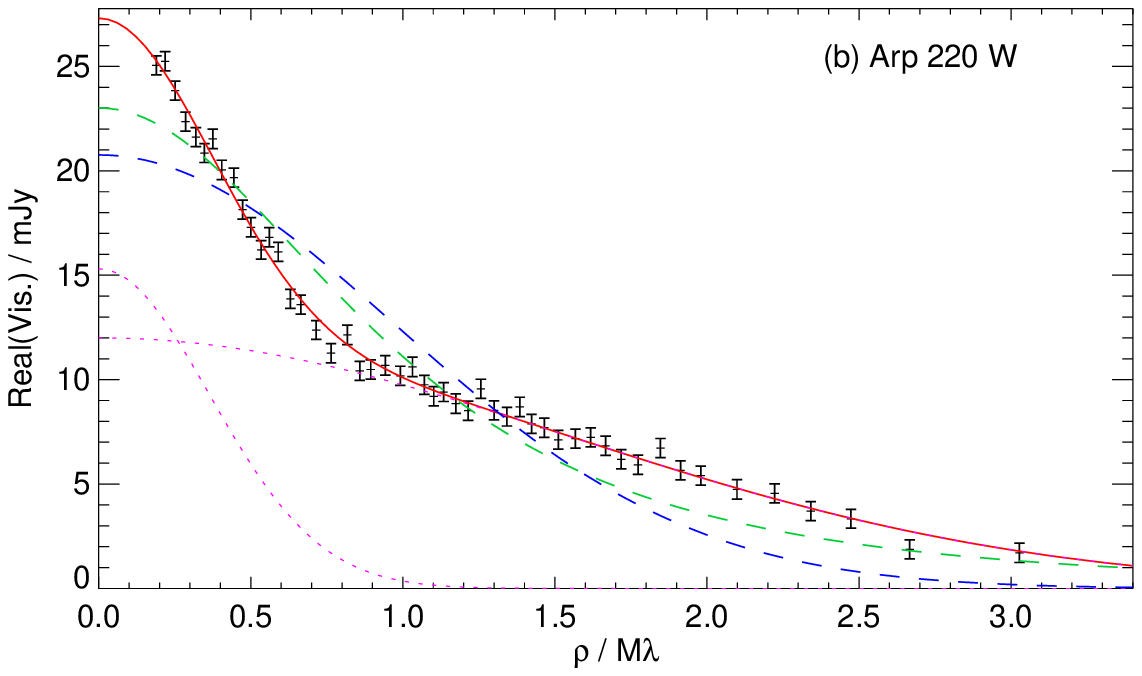}
\caption{ \label{f.B3_r2Gfit}
Arp 220 visibility fitting results for continuum-dominated channels around 112.6 GHz (a segment of \CDCtwo).
Real part of visibilities are plotted as a function of the semi-minor axis of the elliptic coordinate in the \uv\ plane (see text for the fitting procedure).
The data error bars are $\pm 1\sigma$.
The magenta dotted curves are the two Gaussians whose sum, the red curve, best fits the data.
The fitted major axis FWHM ($\theta_{\rm maj}$) at this frequency are the following.
East nucleus:  $\theta_{\rm maj}^{(1)}=74\pm16$ mas and  $\theta_{\rm maj}^{(2)}=381\pm27$ mas.
West nucleus:  $\theta_{\rm maj}^{(1)}=50\pm1$ mas and  $\theta_{\rm maj}^{(2)}=212\pm7$ mas.
For comparison, the blue and green dashed curves show the best fits with a single Gaussian
and an exponential disk, respectively.
}
\end{figure*}

We next fitted each nucleus with restricted two Gaussians (r2G~fit) to see if we need more than one component for each nucleus.
Minimally generalizing the 1G model, the r2G model has for each nucleus two elliptical Gaussians sharing the same center, 
axial ratio, and position angle. 
The shared parameters are fixed to the ones from the 1G fit.
This model naturally defines an elliptic coordinate for each nucleus to plot visibilities as a function of the elliptic radius 
in the same manner as often done for a single axisymmetric source. 
For each nucleus, we first subtract the other nucleus and shift the phase center to the target nucleus, 
then we vector average the visibilities in elliptical annuli and fit the real part of the visibilities.
(In all the other fitting in this section, complex visibilities were simultaneously fitted for the two nuclei without such subtraction.)
The subtraction helps to visualize the fit because otherwise the two nuclei make stripes of 
visibility amplitude and phase in the uv plane. Without radial symmetry, any radial visibility plot will be less informative and harder to compare with models.
We made two iterations to allow a better nucleus subtraction. 
The subtracted models were first from the 1G fit and then from the r2G fit obtained in the previous iteration.
Any subtraction residual causes visibility ripples of 0.2 M$\lambda$ spacing, which corresponds to 1\arcsec. 
They tend to cancel out when averaged in an elliptical annulus larger than 0.2 M$\lambda$. 
Since the cancellation is incomplete in smaller annuli, 
baselines shorter than 0.2 M$\lambda$ were flagged to suppress any subtraction residuals. 
The fitting was made for each of the nine sections of \CDCtwo. 
All resulted in fits similar to an example in Figure~\ref{f.B3_r2Gfit}, visually confirming that a single Gaussian 
is a poor fit for each nucleus while two (restricted) Gaussians can fit the data much better.

This result agrees with our observation at \about250 and 350 GHz  (Sakamoto et al.~in prep.).
It is also in line with the finding of \citest{Barcos-Munoz15} that
33 GHz (9 mm) continuum emission of individual nuclei is better fitted (in the image domain) by an exponential disk than by a Gaussian,
because an r2G model and an exponential disk share a central cusp and slow outer decline.
On the other hand, at 3~mm, neither nucleus is fitted by
a single elliptical exponential disk\footnote{
An axisymmetric exponential disk with a scale length $a$,
\[
	f(x,y) = e^{\left. -\sqrt{x^2+y^2}\right/a},	
\]
has the Fourier transform
\[
	\int_{-\infty}^{\infty} \! \int_{-\infty}^{\infty} 
	f(x,y)
	e^{-i 2 \pi (ux + vy)} \, dx dy
	=
	\frac{2\pi a^2}{\left[1+ 4 \pi^2 a^2 (u^2+v^2) \right]^{3/2}}.
\]
}
as nicely as it is by the r2G model, 
 although the difference is relatively small for Arp 220 E (see  Fig.~\ref{f.B3_r2Gfit}).

\subsection{2G fit} \label{s.visfit.2G}
\begin{figure*}[t]
\epsscale{1.17}
\plottwo{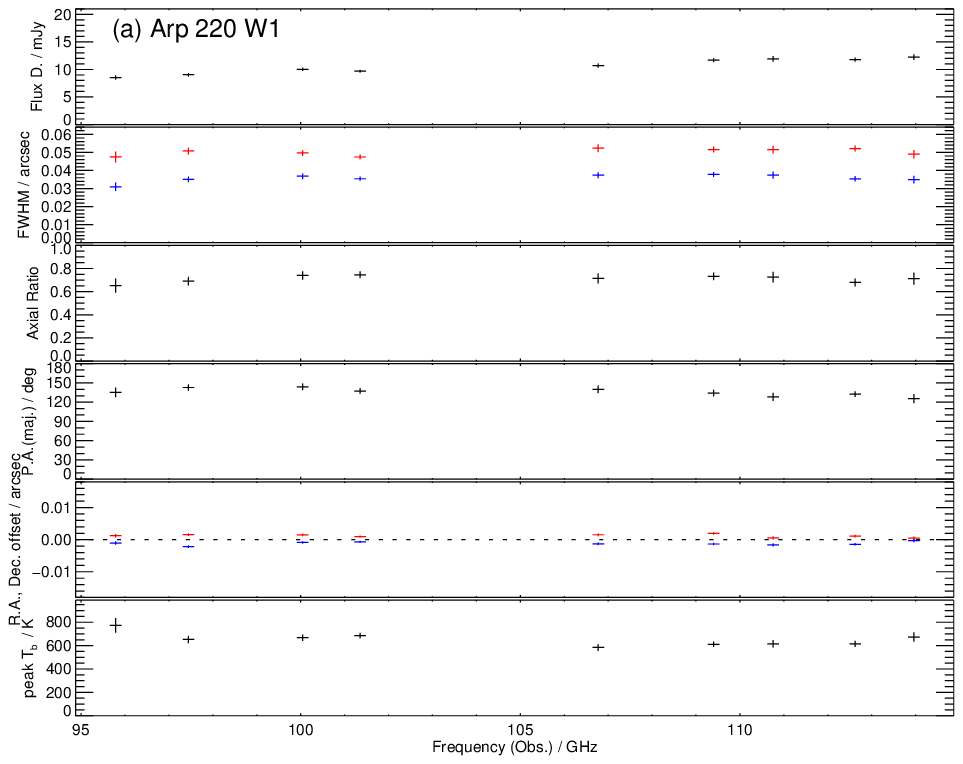}{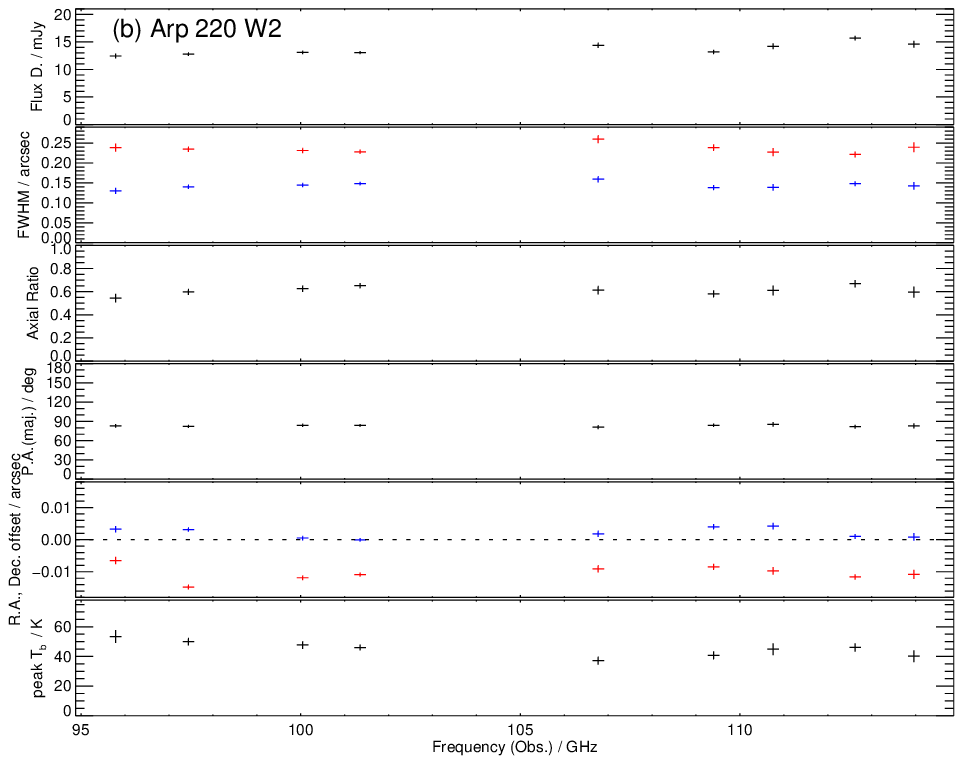}
\plottwo{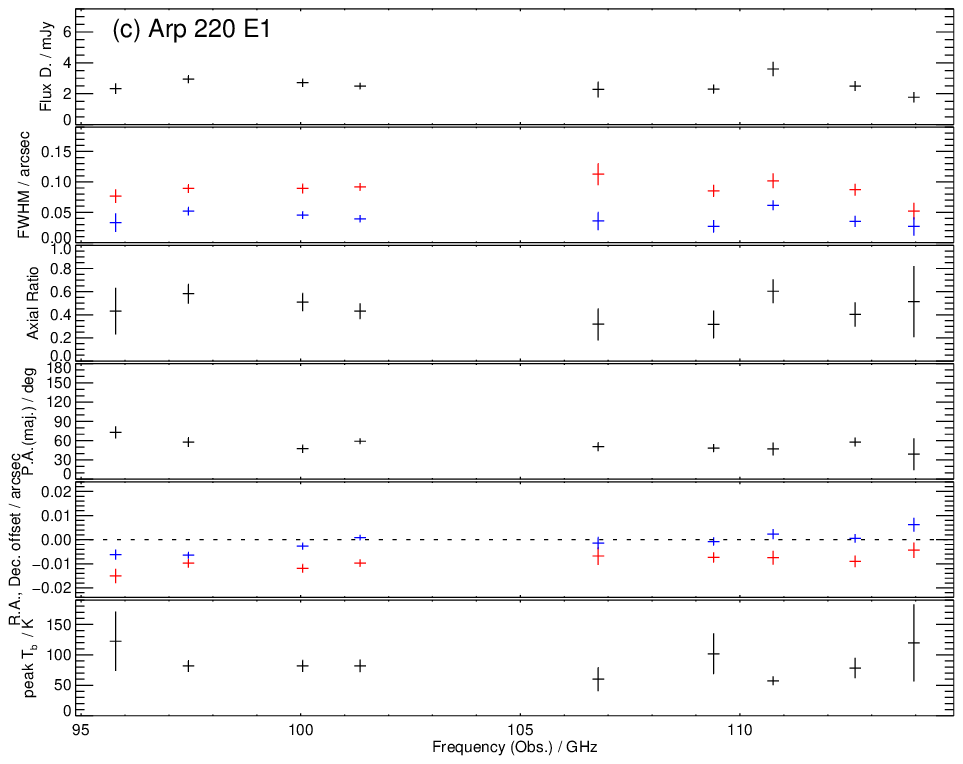}{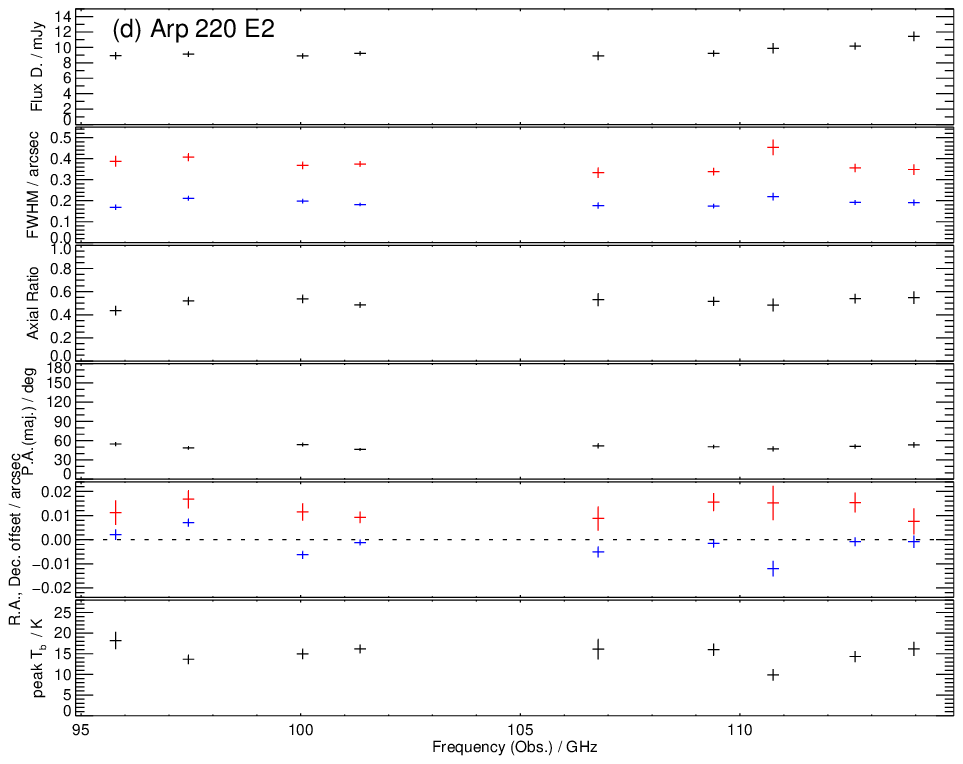}
\caption{ \label{f.B3_2GFullfit}
Arp 220 visibility fitting results for selected continuum-dominated channels (\CDCtwo).
Visibilities were fitted with four elliptical Gaussians, two for each nucleus (2G fit).
The six parameters of each Gaussian and two derived from them (axial ratio and peak brightness) are plotted.
(a) and (b) are for the smaller and larger of the two Gaussians, respectively, to fit the west nucleus.
(c) and (d) are for the east nucleus.
The six sub-panels for each Gaussian show, from top to bottom, 
the source-integrated flux density, 
major (red) and minor (blue) axis FWHM, 
minor-to-major axial ratio,
position angle of the major axis, 
positional offset (red for R.A. and blue for Dec.) from the 1G-fitting position of the nucleus,
and peak Rayleigh-Jeans brightness temperature of the component.
Error bars are $\pm 1\sigma$ and rescaled so that reduced $\chi^2$ will be unity at each frequency.
}
\end{figure*}

After verifying that a single Gaussian provides a poor fit for each nucleus while two Gaussians can do much better, 
we fitted the continuum-dominated visibilities simultaneously using four Gaussians, two for each nucleus, without parameter restrictions (2G fit).
Each Gaussian has six free parameters, one for total flux density, two for position, and three for shape.
The fits easily converged with reduced $\chi^2$ of 1.4--2.1.
Figure \ref{f.B3_2GFullfit} shows that the fitted parameters agree well among the nine \CDCtwo\ segments
even though they are from two observing sessions with independent calibration except for the common flux calibrator model.
This assures that our spatial decomposition is robust and not limited to the r2G model.
Table \ref{t.ParamSummary} lists averages of the derived parameters of individual components.
Denoting the smaller component in each nucleus with a subscript 1 and the larger with 2, 
\Eone\ and \Wone\ have major axis FWHM of 88 mas (36 pc) and 50 mas (21 pc), respectively.
Their peak brightness temperatures are as high as 73 K in \Eone\ and 640 K in \Wone.
They are more than twice smaller in size and warmer in brightness temperature than the single-Gaussian estimates in \citest{Scoville17}.
The larger components \Etwo\ and \Wtwo\ have major axis FWHM 0\farcs37 and 0\farcs23 (about 150, 100 pc) 
and peak brightness temperatures of 15 and 45 K, respectively. 
\Wone\ and \Wtwo\ have their major axes misaligned by about 50\degr\ while
the eastern nucleus is reasonably fit with two Gaussians sharing virtually the same axial ratio and position angle.
In each nucleus the compact and extended components are almost concentric with only 13--22 mas (5--9 pc) offsets between 
their centroids.
It is noteworthy that the spectral index of \Wone, $1.94\pm0.16$, is consistent with that of 
the Rayleigh-Jeans part of optically thick emission, 
although the index can be instead due to a superposition of multiple kinds of emission with different spectral indices.
The compact components \Eone\ and \Wone\ have about 20\% and 47\% of the 3 mm flux densities of the individual nuclei, respectively.
Finally, we verified our 2G fitting by using data before any self-calibration.  
Sizes changed little and within the listed $1 \sigma$ uncertainties. 
The largest change in peak \Tb\ was only 5\% reduction for \Wone\ to $607\pm 14$ K.

\subsection{1G+1E fit}  \label{s.visfit.1G1E}
We made further fitting for parameters of dust emission in the nuclei.
Plasma emission, i.e., synchrotron and free-free emission, dominates at 33 GHz and was fitted well in each nucleus
as an exponential disk \citesp{Barcos-Munoz15}.
We therefore employed two exponential disks having the parameters of the plasma disks (see Table \ref{t.ParamSummary}),
fractional contributions of the plasma and dust emission at 3 mm as we estimate in \S \ref{s.f_dust},
and two Gaussians without fixed parameters. 
Each nucleus is therefore described as an exponential disk of fixed parameters plus a Gaussian (1G+1E fit). 
We refer to the Gaussian dust components in the two nuclei \Eoneprime\ and \Woneprime\ and the plasma disks \Etwoprime\ and \Wtwoprime.
The employed flux contributions of dust emission are \Eoneprime/E = 13\% and \Woneprime/W = 41\% at 3 mm.
This 1G+1E fitting provided only an upper-limit size for the dust component in Arp 220 E 
but the limit is comparable to the size of \Eone\ in the 2G fit.
Because \Eone\ has \about20\% of the 3 mm flux density in the eastern nucleus while \Eone$'$ has 13\%, we see that \Eone \about \Eoneprime.
We obtained consistent results about Arp 220 W over the \CDCtwo\ segments
and the averaged parameters are in Table~\ref{t.ParamSummary}.
The dust component \Woneprime\ has a spectral index $3.57 \pm 0.21$ as a direct consequence of our assumed
spectral index of $\alphad=3.8$ for the dust emission. 
Except for the spectral index, the parameters of \Woneprime\ are close to those of the 
component \Wone\ in our 2G fit including the flux density fraction of \Wone/W = 47\%, hence $\Wone \approx \Wone'$.
Combining the results for the two nuclei, 
the compact components in our 2G fit represent mostly dust emission peaking at the centers of the two nuclei 
while the larger components in individual nuclei correspond to the exponential (circum)nuclear disks with dominant plasma emission. 
We are going to further discuss the decomposition in \S \ref{s.discussion}.

\clearpage 
\section{Images} \label{s.images}
We made continuum images from the continuum-dominated channels \CDCone\ introduced in Section \ref{s.spectra}.
Multi-frequency synthesis was made with two terms (i.e., the spectrum at each position is assumed to follow
a power law) and using the robust weighting \citep{Briggs95}.  

\begin{figure}[htb]
\epsscale{0.6}
\plotone{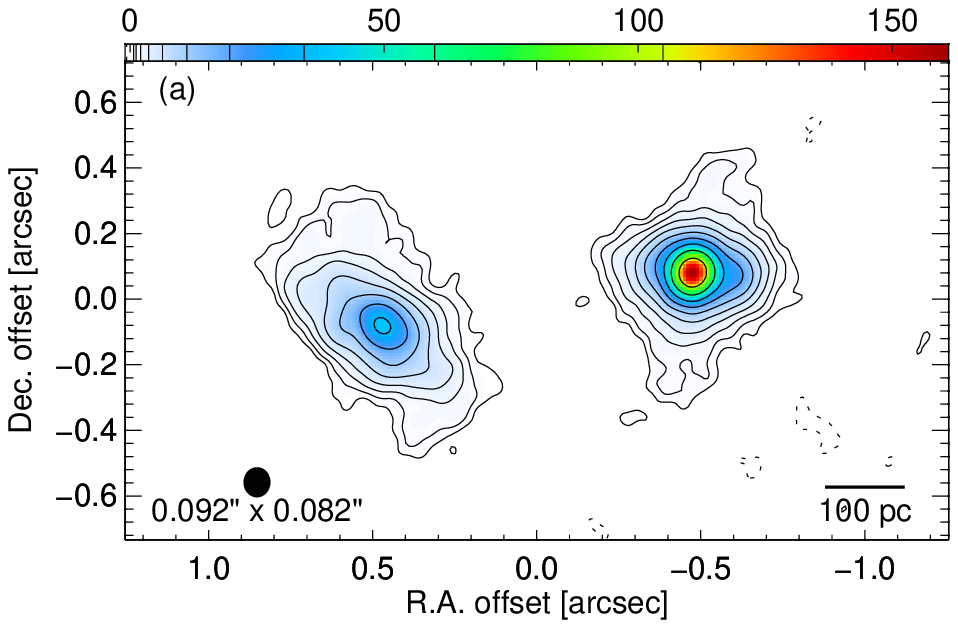}\\
\plotone{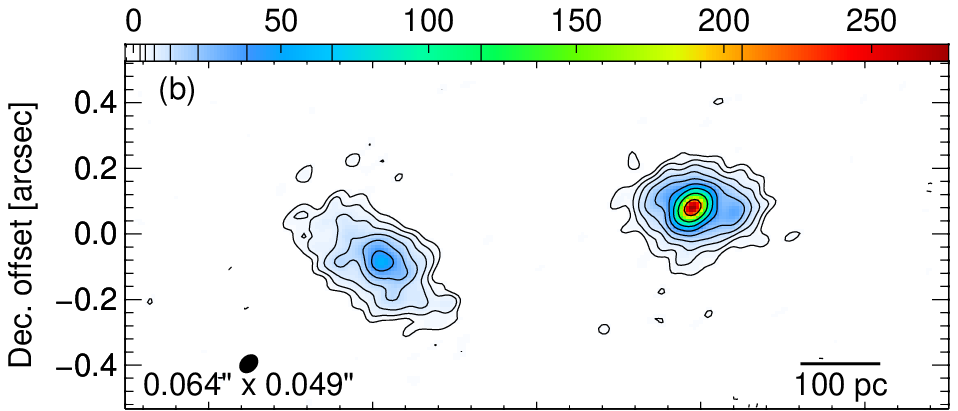}\\
\plotone{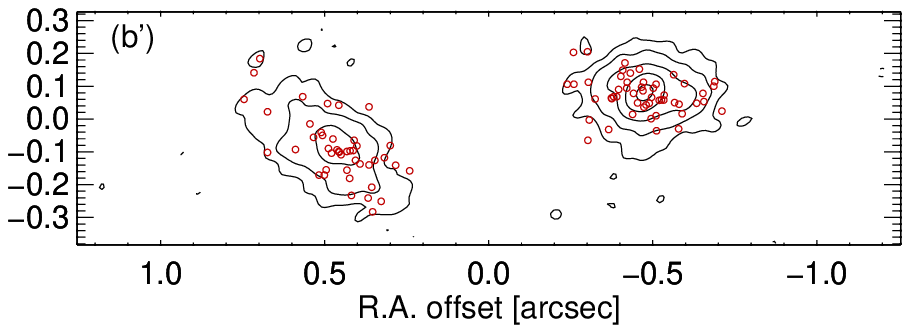}\\
\plotone{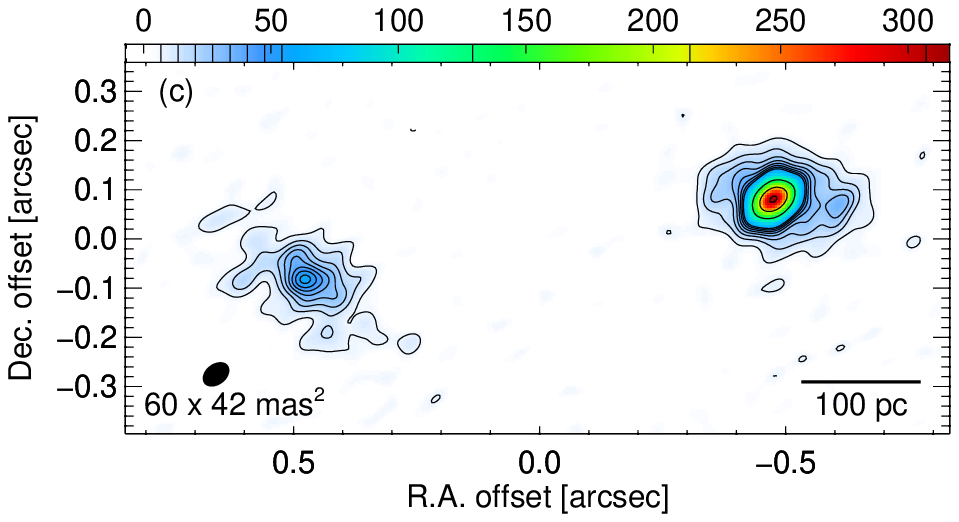}

\caption{ \label{f.B3contIm}
Arp 220 continuum images at 104.1 GHz (2.9 mm). 
Each panel has the FWHM beam size at the bottom-left corner.
The offset coordinates are from the midpoint of the two nuclei.
The intensity unit of the color bars is kelvin (Rayleigh-Jeans brightness temperature). 
(a) Imaged with {\tt robust=2}, i.e., the natural weighting. The $n$-th contour is at $\pm3n^{1.75}\sigma$ where $\sigma=0.23$ K 
(15 \micro Jy \perbeam). 
The peak intensity is 161 K (10.8 mJy \perbeam).
Negative contours are dashed.
(b) Imaged with {\tt robust=0}. The $n$-th contour is at $\pm3n^{1.75}\sigma$ where $\sigma=0.78$ K. 
The peak intensity is 276 K.
(b') The same image as (b) with every other contours. 
Over-plotted are compact ($ < 2$ pc) radio VLBI continuum sources \citep{Varenius17}.
(c) Imaged with {\tt robust=$-$1}.
The $n$-th contours are at $\pm 3n \sigma$ ($1\leq n\leq8$) 
and $3\times8(n-7)^{1.25}\sigma$ ($8 \leq n$), where $\sigma=2.3$ K.
The peak intensity is 57 K in Arp 220 E and 316 K in Arp 220 W.
}
\vspace{13mm}
\end{figure}

%
Figure \ref{f.B3contIm} shows the 3 mm continuum images made with three different robust parameters. 
Dominant features in all are a bright compact source at the center of Arp 220 W (\Wone\ in our 2G fit)
and larger elongated structures in the position angles of about 50\degr\ in E and 80\degr\ in W (\Etwo\ and \Wtwo, respectively).
Arp 220 E also has a compact central component (\Eone) that is easier to see in the panel (c)
with the highest resolution and equal-step contours at lower levels.
The overall structure is consistent with our visibility fitting results.
The peak intensities in the maps are lower than our fitted values as expected from beam dilution.

Looking for features that could not be captured with our simple models,
we find in Fig.~\ref{f.B3contIm}(a) that the western nucleus has faint bipolar features that extend from the central region along
P.A. $\sim 170$\degr\ and perpendicular to the major axis of \Wtwo. 
This feature is below the lowest contour in the continuum image of \citest{Scoville17}, which corresponds to our third contour.
The eastern nucleus also shows faint features in the panel~(a) around the tips of its major axis, 
extending in the opposite directions like an integral sign. 
This is also seen at 33 GHz \citesp{Barcos-Munoz15}.
In the panels~(b) and (c), we see that the extended component \Etwo\ starts to be resolved to multiple sources.
This may be also the case for \Wtwo\ in the panel~(c).

Fig.~\ref{f.B3contIm}(b$^\prime$) compares the 3 mm continuum emission with the distribution of 
compact ($ < 2$ pc) radio (18--2 cm) sources that were observed over 1994--2014 with VLBI \citep[and references therein]{Varenius17}.
These sources have been almost entirely attributed to supernovae, i.e., radio supernovae and young supernova remnants
\citep{Smith98, Lonsdale06, Parra07,Batejat11,Varenius17}.
The VLBI source distribution is similar to the 3 mm emission distribution in both nuclei, mainly following
the larger \Etwo\ and \Wtwo\ components.
The similarity was also seen by \citest{Barcos-Munoz15} with 33 GHz plasma emission
at a comparable resolution of 0\farcs07.

\clearpage 
\section{Further Analysis and Discussion} \label{s.discussion}

\subsection{Nuclear Disks and Outflow from Arp 220 W} \label{s.disks_outflow}
We ascribe the larger components (\Etwo\ and \Wtwo) in our two-Gaussian (2G) model
to the nuclear disks of Arp 220 that rotate around individual nuclei \citep{Sakamoto99}. 
We attribute the elongation of these components on the sky to the disk inclinations and
suggest that the faint feature perpendicular to the major axis of the western nuclear disk is 
a bipolar outflow.

The (counter-)rotating nuclear disks have been seen through velocity gradients along their major axes 
in high-resolution line imaging \citep{Sakamoto99,Sakamoto08,Scoville15,Scoville17}. 
If the disks are thin then their inclinations can be calculated from their axial ratios 
to be 60\degr\ for \Etwo\ and 52\degr\ for \Wtwo, which agree with the estimates of \citest{Barcos-Munoz15} for \Etwoprime\ and \Wtwoprime.
At least our inclination for the western nuclear disk is a lower limit because the outflow emission biases the measured axial ratio
and the outflow-driving activity in the disk should give it a thickness.
We thus estimate $i \approx 60\arcdeg$ for both disks. 
An apparent minor-to-major axis ratio of 0.62 for the western nuclear disk translates to an inclination 
in the range of 60\degr\ $\pm$ 10\degr\ for an oblate spheroid having a hight-to-radius ratio up to 0.55.
On the western nuclear disk we therefore differ with \citest{Scoville17} who,
without seeing the faint bipolar component,
modeled it as a thin face-on disk of $i\approx 30\degr$.
The CO velocity field in their Fig.~4 does suggest a nearly face-on configuration {\it if} 
the gas distribution is two-dimensional. 
However, a similar velocity field can be due to a three-dimensional bipolar outflow seen from the side \citep{Seaquist01,Walter02}. 
Therefore, while it remains to be seen whether the CO data cube can be reasonably fitted with 
a three-dimensional outflow model, at least a projection of the cube (i.e., the mean velocity map) is 
consistent with a nearly edge-on outflow. 
The large inclination of the western nuclear disk explains the alignment of the supernova features 
(Fig.~\ref{f.B3contIm}b$'$), 
which posed difficulty in \citest{Scoville17}'s face-on disk model because any alignment at the time of star formation 
would be erased in the differentially-rotating disk by the time of supernova explosions.
The large inclination of the western nuclear disk had been the preferred model since \citet{Scoville98} inferred it
from near infrared imaging with the Hubble Space Telescope.

Molecular outflow from individual nuclei of Arp 220, in particular from W, has been known from 
P-Cygni line profiles and blueshifted line absorption 
\citep{Sakamoto09,Rangwala11,Gonzalez-Alfonso12,Veilleux13,Tunnard15,Martin16,Barcos-Munoz16,Zschaechner16},
which is also evident in the high resolution spectrum in Fig.~\ref{f.B3spec} for the western nucleus.
\citet{Sakamoto09} also noted that OH masers observed by \citet{Rovilos03} show a bipolar distribution
around the western nucleus along its disk minor axis in the north-south direction, as one would expect for an outflow from the nuclear disk.
\citet{Tunnard15} proposed this to be the actual outflow configuration
on the basis of their finding that SiO(6--5) at the systemic velocity is in absorption to the south and emission to the north of the nucleus.
\citet{Varenius16} further detected north-south extension of 150 MHz emission around the western nucleus 
at \about0\farcs5 resolution and attributed it to a bipolar outflow of the same configuration.
\citet{Barcos-Munoz16} found that a spectral index map between 33 and 92 GHz at \about0\farcs08 resolution
has a distinct positive-value region across the western nuclear disk along P.A. \about\ 153\arcdeg.
Our observation of a bipolar structure in 3 mm continuum is consistent 
with these observations and corroborates the outflow from Arp 220 along P.A. \about\ 170\arcdeg.
{Further evidence supporting this outflow will be presented in Barcos-Mu\~{n}oz et al (in preparation).}

%
In light of the revived picture of a highly inclined disk plus a bipolar outflow for the western nucleus, 
we notice that the CO maps of the nucleus in \citest{Scoville17} (their Fig.~4) show a bipolar feature corresponding 
to that in the 3 mm continuum.
Its southern part is more prominent and blueshifted.
\citet{Tunnard15} already proposed that the southern side of the bipolar outflow is approaching to us, i.e., blueshifted.
For this velocity structure, the southern side of the western nuclear disk must be its far side if the outflow axis is normal to the disk.
This is at odds with the near infrared observation that the southern side has much larger extinction and 
should be the near side of the nuclear disk \citep{Scoville98}.
Therefore, there remain unsolved problems about the detailed configuration of the outflow 
and the inner structure of the western nuclear disk.
It may be that the outflow axis is not perpendicular to the disk but only in the plane containing the disk rotation axis and our sight line,
or the disk may be warped.
Regarding the outflow driver, both nuclear disks at least have vigorous star formation as traced by the VLBI supernovae.
In addition, the core component \Wone, with its very high luminosity surface density (see \S \ref{s.WoneLuminositySource}), 
probably makes a large or even dominant contribution to drive the outflow.
The outflow has oblique angles with respect to the major and minor axes of \Wone\ (and \Woneprime) while it is 
orthogonal to the \Wtwo\ major axis.
This implies a role of the nuclear disk for outflow collimation but, as noted above, three dimensional configuration of the system 
is necessary to verify this.
On Arp 220 E, while blueshifted absorption was also detected toward it, the lack of a bipolar feature at 3 mm in this nucleus despite 
both nuclear disks apparently having large inclinations suggests that the outflow from the eastern nucleus has
a less mass or a less 3 mm emissivity or both.

\subsection{Fractional Contribution of Dust Emission} \label{s.f_dust}

We estimate from spectral indices that the fractional contribution of dust emission to the 3 mm continuum is
$\fdust \approx$ 13\% and 41\% for Arp 220 E and W, respectively.
If dust emission with a spectral index \alphad\ makes a fractional contribution \fdust\ to the observed
flux density and the rest of the emission from plasma, which is the sum of synchrotron and free-free emission, has 
a spectral index \alphap\ then the total emission 
has a spectral index of $\alpha = \fdust \alphad + (1 - \fdust) \alphap$.\footnote{
A spectrum that consists of power law spectra having spectral indices $\alpha_i$ and fractional contributions $w_i$
at a reference frequency $\nu_0$, 
\[ 
S_\nu = S_{\nu_0} \sum_i w_i \left( \frac{\nu}{\nu_0} \right)^{\alpha_i},
\quad \mbox{where } \sum_i w_i = 1,
\]
has the spectral index
\[ 
\alpha = \left. \frac{d\, \log S_\nu}{d\, \log \nu} \right|_{\nu = \nu_0} 
                 = \left[ \frac{\nu}{S_\nu}  \frac{d\, S_\nu}{d\, \nu}   \right]_{\nu = \nu_0}
                 = \sum_i w_i \alpha_i
\]
at the reference frequency.
} 
We use $\alphad = 3.8$ for optically thin dust emission using the dust emissivity index $\beta$ of $1.8\pm0.1$ in the Galactic plane \citep{Planck21}
and $\alphap = -0.59 \pm 0.08$ for E and $-0.61\pm 0.07$ for W from the 6--33 GHz measurements by \citest{Barcos-Munoz15}.
With the overall 3 mm spectral indices that we estimated from Fig.~\ref{f.B3spec}, we obtain 
$\fdust = 0.13 \pm 0.04 $ and $0.41 \pm 0.05$ for Arp 220 E and W, respectively. 
The errors do not include the effect of any variation of the plasma spectral index between 6--33 GHz and \about100 GHz.
Although \alphap\ likely increases at higher frequencies as the fractional contribution of free-free emission increases,
its effect to \fdust\ should be small because \citest{Barcos-Munoz15} found from cm-wave data that synchrotron emission dominates at 33 GHz
(and attributed the weakness of free-free emission to dust absorption of ionizing photons). 
For example, if synchrotron emission has a constant spectral index $-0.7$ and free-free emission at $-0.1$ then the fractional contribution of
synchrotron to the 33 GHz continuum having $\alpha = -0.6$ should be 5/6 = 83\%. 
The synchrotron fraction decreases only slightly to 71\% and \alphap\ increases only slightly to $-0.53$ at 104 GHz. 
For  this \alphap\ the dust contribution to 3 mm continuum  will decrease only by 0.01 for each nucleus.
Likewise, if a part of the dust emission is saturated (i.e., opacity $\gtrsim 1$) at 3 mm or the dust $\beta$ is smaller than assumed
(e.g., median $\beta$ is 1.6 in single-temperature fits for a sample of ULIRGs \citep{Clements10}) then \fdust\ increases slightly,
only by 0.01 for E and 0.03 for W for \alphad=3.5.
It is notable that for each nucleus the estimated fraction of dust emission broadly agrees with the flux-density fraction of the compact component 
in our 2G fit, 20\% for E and 47\% for W (\S \ref{s.visfit.2G}).

One could also estimate dust emission by subtracting the plasma component from the observed total emission. 
Arp 220 E and W should have 104.1 GHz flux densities of $15.1 \pm 2.3$ and $16.4 \pm 2.4$ mJy, respectively,
from synchrotron and free-free emission if we 
extrapolate the 33 GHz flux densities with the 6--33 GHz spectral indices in \citest{Barcos-Munoz15}.  
The errors include 12\% absolute flux calibration uncertainty at 33 GHz.
The ALMA total flux densities at the same frequency are estimated to be $11.9\pm0.9$ and $23.9\pm1.3$ mJy, for E and W respectively, 
from the 2G fit assuming a power-law among the nine frequency segments; the uncertainties include 5\% error in absolute flux scale.
Nominally, \fdust\ is calculated from these flux densities to be $-0.27$ and 0.31 in Arp 220 E and W, respectively.
The unphysical negative fraction may be simply because the denominator or numerator or both are in error;
the fraction could be zero if both are in error by 1$\sigma$ from the estimates above.
Other possible sources for error include the flux calibration both at VLA and ALMA, missing flux in ALMA data
that may be larger for a more extended eastern nucleus, and change in spectral index of the plasma emission between 33 and 100 GHz. 
Because the previous method to estimate \fdust\ from the spectral indices alone is not affected by the first two of these errors, 
we adopt the \fdust\ estimates in the previous paragraph.
We note that \citest{Scoville17}, despite using this direct subtraction method, could obtain \fdust\ of 0.11 and 0.45 respectively 
for E and W at 112.6 GHz.
Our adopted estimates agree with theirs.

\subsection{Distribution of Dust Emission in the Nuclei} \label{s.dustEmission}
We can estimate the spatial distribution of dust emission in the nuclei by subtracting plasma emission. 
For the latter emission we use the exponential disks that \citest{Barcos-Munoz15} fitted to the Arp 220 nuclei at 33 GHz.
We adopt their deconvolved shapes, place the plasma disks in our 1G fit positions, and scale the model flux densities 
to be consistent with the ALMA flux densities of the nuclei multiplied by their adopted $(1-\fdust)$.
Within the 100 GHz band we assume that the plasma disks retain their 33 GHz spectral indices.

\begin{figure}[htb]
\epsscale{0.6}
\plotone{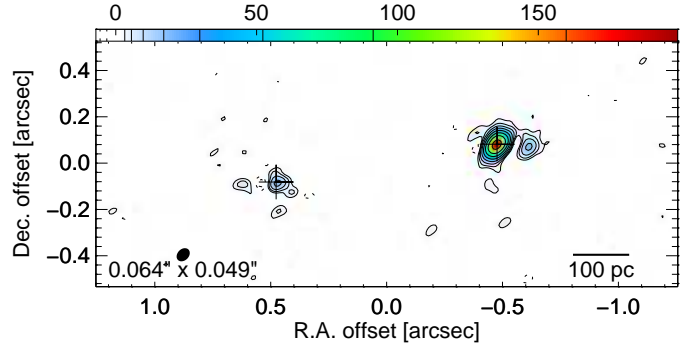}\\
\caption{ \label{f.B3dustIm}
Arp 220 dust continuum emission at 104.1 GHz (2.9 mm).
It is estimated by subtracting plasma emission in the forms of elliptical exponential disks measured at 33 GHz. 
This image was made with {\tt robust=0} after the subtraction in the \uv\ domain. 
The FWHM beam size is at the bottom-left corner.
The offset coordinates are from the midpoint of the two nuclei.
The crosses are at the continuum 1G-fit positions in Table \ref{t.ParamSummary}.
The $n$-th contour is at $\pm3n^{1.75}\sigma$ where $\sigma=1.06$ K 
(30 \micro Jy \perbeam).  Negative contours are dashed.
The peak intensity is 20 K in Arp 220 E and 199 K in Arp 220 W.
The intensity unit of the color bar is kelvin (Rayleigh-Jeans brightness temperature). 
}
\end{figure}

Figure~\ref{f.B3dustIm} is a dust emission map made using all \CDCtwo\ data. 
The dust emission in each nucleus has the strongest peak at the center and also has weaker peaks around it.
The central concentration of the dust emission is much more pronounced than that of the VLBI sources in both nuclei and in particular in the western nucleus. 
This compactness of dust emission compared to VLBI supernova distribution in Arp 220 W was already noted 
at 0.86 mm \citep{Sakamoto08}.
The central peak of 3 mm dust emission is much stronger in Arp 220 W than in Arp 220 E. 
This partly corresponds to the larger fractional contribution of dust emission in Arp 220 W (\about41\% at 3 mm) than in
Arp 220 E (\about13\%).
The weaker peaks around the central ones may be dust emission from the nuclear disks but some of them can be 
residual plasma emission because we only subtracted its parametrized approximation.
We also caution about an underlying assumption in this subtraction that all emission is optically thin.
If the dust emission toward the center of a nucleus is optically thick then it is unnecessary to subtract plasma emission
from behind the dust photosphere because such emission does not reach us in the first place.
It is therefore possible that the dust emission peaks are more pronounced than seen in Fig.~\ref{f.B3dustIm}
at the centers of the two nuclei, in particular the western nucleus where the spectral index of \Wone\ is compatible with optically thick emission.

%
We already obtained the parameters of the dust emission through the visibility fitting with the 1G+1E model (\S \ref{s.visfit.1G1E}).
In the fitting, the parameters of the model exponential disks were fixed to those of the plasma emission.
The results for the Gaussian (dust) components are consistent with our image-domain estimate for dust emission 
regarding the presence of compact peaks at the centers of the two nuclei.
The caution about the possible over-subtraction toward the center of the western nucleus applies to this visibility fitting too.
The peak brightness temperature of the dust emission at the center of the western nucleus, $\gtrsim 500$ K, is more than twice higher than
any previous measurements of dust brightness temperature in Arp 220. 
For example, 
\citet[\about90 K]{DownesEckart07}, 
\citet[\about160 K]{Sakamoto08}, 
\citet[\about200 K]{Wilson14}, 
and 
\citet[\about120 K]{Scoville17} all reported deconvolved peak 
brightness temperatures of dust continuum only up to 200 K with single-Gaussian deconvolved FWHM of \about0\farcs1--0\farcs2.
There have been, however, pieces of spectroscopic evidence for much warmer dust and molecular gas in the galaxy 
\citep[e.g.,][]{Martin11,Rangwala11,Gonzalez-Alfonso12,Gonzalez-Alfonso13}.

\subsection{Bolometric Luminosity of the Compact Core in the Western Nucleus}  \label{s.Lbol}
The bolometric luminosity of the compact core found in the western nucleus can be estimated as follows for its thermal dust emission.
A geometrically-thin and optically-thick disk with a radial temperature described by a Gaussian falloff has
a bolometric luminosity of 
\begin{eqnarray} \label{eq.lbol_gaussian_disk}
	\Lbol 
	& = & 
	2 \int_{-\infty}^{\infty} \! \int_{-\infty}^{\infty} 
	\sigma 
	\left[ 
	T_{\rm p} 
	\exp\left(- \log 2 \; \frac{x^2+y^2}{r_{\rm maj}^2} \right) 
	\right]^4 
	dx \, dy	
	\nonumber \\
	& = &
	\frac{\pi \rmaj^2 \sigma \Tp^4 }{2 \log 2}
	\nonumber \\
	& = &
	3.2\times 10^9 
	\left(
		\frac{r_{\rm maj}}{\mbox{10 pc}}
	\right)^2
	\left(
		\frac{\Tp}{\mbox{100 K}}
	\right)^4	\Lsol,
\end{eqnarray}
where the factor of 2 before the integral is for the two faces of the disk, 
$\sigma$ is the Stefan-Boltzmann constant,
\Tp\ is the peak dust temperature,
and
\rmaj\ is (major-axis FWHM)/2 of the Gaussian temperature distribution in linear scale.
For \Woneprime, we obtain \Lbol = $(3.3 \pm 0.8) \times 10^{12}$ \Lsol\ 
from \rmaj=$11.5 \pm 0.3$ pc and \Tp = $529 \pm 20$  K as well as 5\% absolute flux-scale uncertainty.
Here brightness temperature of the dust emission is equated with dust physical temperature. 
This \Lbol\ is four times larger than the one for the western nucleus by \citet{Wilson14}, who used the same formula and 0.43 mm data
(\Tp\ = 197 K and  \rmaj\ = 41 pc scaled to our adopted distance),
and is 1.7 times larger than the \Lir\ of Arp 220.

There are cautions for and limitations in the luminosity estimate and comparison above.
First, our subtraction of plasma emission had an assumption that the dust (as well as plasma) emission is optically thin at 3 mm
while the calculation above assumes an optically thick emitter.
Even if the dust emission around the peak has a 3 mm opacity below unity it cannot be much below
because dust cannot be hotter than its sublimation temperature; 
one obtains $\tau_{\rm 3mm} > 0.25$ from $\max(\Tb)/\tau_{\rm 3mm} \approx \Tdust < 2000 $ K.
The dust is therefore expected to become optically thick at shorter wavelengths not far from 3 mm 
because of the wavelength-dependent opacity coefficient. 
Observations at shorter wavelengths also suggest so \citep{Sakamoto08,Wilson14}.
In such a situation, the bolometric luminosity of the \Woneprime\ component can be higher than the calculation above 
because at shorter wavelengths where most of the luminosity is radiated the nucleus can radiate at a higher brightness temperature than at 3 mm,
at \Tdust\ instead of $(1 - e^{-\tau})\Tdust$.
To rectify, we can adopt an assumption that the 3 mm continuum toward the center of Arp 220 W is optically thick, which
is consistent with the spectral index of \Wone\ in the 2G fit. 
Using the \Wone\ parameters, the bolometric luminosity of the core is calculated to be $(5.7\pm1.2)\times 10^{12}$ \Lsol.
Second, the total luminosity from Eq.~(\ref{eq.lbol_gaussian_disk}) integrates direction-dependent radiation from the disk 
over the entire directions whereas the observational source luminosity \Lir\ is based on our measurements from a single direction
and is calculated assuming isotropy. 
The latter luminosity can be biased for a disk-like source with anisotropic radiation.
Most of the bolometric luminosity of Arp 220 is observed at mid-to-far infrared wavelengths around 50 \micron.
Hence most of the luminosity from the 500 K core is absorbed and re-radiated before reaching us, presumably
in large part by the nuclear disk \Wtwo. 
Because we look at \Wtwo\ (as well as \Etwo) from the side, i.e., from directions with less flux, 
\Lir\ of Arp 220 may well be underestimated.
Third, we assumed that each nucleus is an axisymmetric disk to derive its inclination and luminosity, but this may not be valid. 
Removing the assumption, the lowest disk luminosity can be obtained by
replacing $r_{\rm maj}^2$ with $r_{\rm maj}r_{\rm min}$ in Eq.~(\ref{eq.lbol_gaussian_disk}). 
It is $(1.9 \pm 0.5) \times 10^{12} \Lsol$ for \Woneprime, which agrees with the \Lir\ of Arp 220.
This is the limiting case in which \Woneprime\ is an oval-shaped disk observed face-on.
Fourth, the true distribution of brightness temperature in \Wone\ as well as \Woneprime\ may not be Gaussian and may be more flat-topped.
Or it may be that the 33 GHz emission has a weak central cusp in addition to the exponential disk but it was missed in the 
observations at 0\farcs07 resolution. If it were due to opaque free-free emission then it can become significant at 100 GHz.
Because of the $\propto T^4$ dependence, \Lbol\  in these cases would be smaller than the calculations above.

To summarize, the bolometric luminosity of the dust thermal emission from the central component in Arp 220 W is estimated to be
\about$10^{12.5}$\Lsol\ with at least $\pm$0.2 dex uncertainty due to various assumptions. 
It is as large as most of the bolometric luminosity of Arp 220 and 
may even exceed the  \Lir\ of Arp 220 estimated from our vantage point.
Higher resolution data will help improve the luminosity estimate by better constraining the shape and temperature of the core.
More frequency coverage will also help to better extract dust thermal emission from the mixture of dust continuum, synchrotron, free-free, and line emission at millimeter and submillimeter wavelengths.
In passing we add that the dust temperatures in \Eoneprime\ and the two nuclear disks are very likely 
higher than their low brightness temperatures in 3 mm dust emission because dust is probably optically thin in these components.
If so they may also have considerable luminosities that will be better constrained with observations at shorter wavelengths.

\subsection{Luminosity Source and Evolution  of the Western Nucleus} \label{s.WoneLuminositySource}
The high peak intensity of continuum emission at the center of Arp 220 W constrains the luminosity source,
with the caveats in the preceding section.
The peak brightness temperature of dust thermal emission,  $\Tp = 5.3 \times 10^2$ K for the \Woneprime\ component, 
translates to a peak luminosity surface density of $\sigma \Tp^4 = 1.1\times 10^{16} $ \Lsol\persquarekpc.
Since we do not know fine details of the spatial distribution of the continuum emission, more robust
than the peak values are the means in the half-light diameter (= FWHM for a Gaussian),
namely the mean brightness temperature $\langle \Tb \rangle_{1/2} = 3.8 \times 10^2 $ K 
and mean luminosity surface density $\sigma (\langle \Tb \rangle_{1/2})^4 = 3 \times 10^{15} $ \Lsol\persquarekpc\
in the central 23 pc.
For comparison, \citet{Soifer03} obtained from mid-IR observations of three Seyfert nuclei 
the surface brightnesses of $(1-5)\times 10^{14}$ \Lsol\persquarekpc\ at similar linear scales (10--30 pc).
The surface brightnesses of infrared-luminous starburst galaxies and Galactic \ion{H}{2} regions 
are typically an order of magnitude (or more) below these Seyfert values \citep{Soifer01,Evans03}.
\citet{Barcos-Munoz17} estimated luminosity surface densities in the 33 GHz half-light radii (30 pc to 1.7 kpc) for 22 local 
ultra/luminous infrared galaxies including those with AGNs.  
Their maximum value is $1\times 10^{14}$ \Lsol\persquarekpc\ and the mode  is at $1\times 10^{13}$ \Lsol\persquarekpc. 
This agrees with earlier analysis by \citet{Thompson05} who not only showed statistics on IR luminous galaxies but
derived $10^{13}$ \Lsol\persquarekpc\ as a characteristic value for warm starbursts ($T < 200$ K) constrained by radiation pressure on dust.
\citet{Soifer03} noted that super star clusters (SSCs) can have as high a luminosity surface density at pc scale 
as the Seyfert nuclei do at a few 10 pc scale. 
For example, the most luminous SSC in the dwarf galaxy NGC 5253 has an age \about 1 Myr,  FWHM size \about 1 pc, 
mass \about $10^{5.5} \Msun$ \citep{TurnerBeck04,Calzetti15}
and hence a luminosity \about$10^{8.5} \Lsun$ and luminosity surface density \about $10^{14.5}$ \Lsol\persquarekpc.
The Arches cluster in the center of our Galaxy also has almost the same luminosity surface density \citep{Lang05}.
From this comparison, empirically speaking, a luminous AGN is a plausible source of luminosity for the central core $\Wone \approx \Woneprime$ 
in the western nucleus. 
Obviously, an empirical argument cannot rule out the possibility that by far the most intense starburst in the local universe is 
at the center of Arp 220 W.
Our new constraint for the case where star formation dominates the luminosity is that
the starburst in the central 20 pc must be equivalent to several thousands of massive young SSCs.
Tighter constraints on the luminosity source are expected from further ALMA observations.

The center of Arp 220 W likely has a column density to make 3 mm dust opacity on the order of unity
and hence a mass surface density of \about$10^6$ \Msol \persquarepc\ (\citest{Scoville17}; \citest{Barcos-Munoz15}; and \S \ref{s.Lbol}). 
The central 20 pc of Arp 220 W is therefore at or above the Eddington limit for dust
\citep[see eq. (7)]{AT11}. The bipolar outflow from Arp 220 W is therefore expected.
The nucleus has been actively forming stars in the nuclear disk, feeding either a luminous AGN or exceptionally intense starburst at the center, 
and blowing out dust and gas at the same time. The nucleus is certainly in a phase of rapid evolution.

\section{Conclusions}
We have analyzed ALMA high-resolution data of Arp 220 at \about3 mm wavelengths.
We spatially resolved continuum structure of the individual nuclei and decomposed the nuclei to plasma and dust emission.
The dust emission was then used to characterize the luminosity source in the western nucleus.
Our major observations are the following.

1. Both nuclei are found to have at least two structural components at 3 mm.
Two concentric components such as Gaussians or a Gaussian and an exponential disk provide reasonable fits to
the observed visibilities.

2. The larger components in the two-Gaussian fit have FWHM sizes \about0\farcs23--0\farcs37  (100--150 pc) and axial ratios \about2.
They match in shape and extent the distributions of supernova features seen with VLBI.
We identify them to the starburst nuclear disks rotating around individual nuclei with inclinations $\approx 60\degr$.

3. The smaller components in the two-Gaussian fit contribute about 20\% and 50\% of the 3 mm
continuum flux densities of the eastern and western nuclei, respectively.
They have FWHM sizes of 20--40 pc and peak brightness temperatures 70--640 K, 
which are more than twice smaller and brighter than in the previous single-Gaussian fit of the same data.

4. The 3 mm continuum spectral slopes are flat ($S_\nu \propto \nu^{0.0 \pm0.2}$) and positive ($\propto \nu^{1.2\pm0.2}$) 
at the eastern and western nuclei, respectively.
Combining them with 33 GHz data of plasma emission, we estimate 
that dust emits about 13\% and 41\% of the 3 mm continuum in the eastern and western nuclei, respectively.
The dust emission is found centrally concentrated in both nuclei.
These central cores of dust emission correspond to the compact components in our two-component fitting.

5. The dust-continuum core of the western nucleus is estimated to have a peak brightness temperature of \about530 K 
and major axis FWHM of about 20 pc after subtracting plasma emission.
Assuming a dust disk, its bolometric luminosity can be as large as
\about$10^{12.5}$ \Lsun\ or at least a large fraction of the total luminosity of Arp 220. 
Its luminosity surface density is on the order of $10^{15.5}$ \Lsol\persquarekpc\ in 20 pc scale. 
This is about an order of magnitude higher than observed toward Seyfert nuclei at the same scale 
and super-star clusters at pc scale.
This comparison favors the presence of a luminous AGN on empirical grounds,  but we stress
the uncertainties still in the data interpretation, the inherent limitation of the empirical argument,
and the need for further observational and theoretical constraints.

6. The western nucleus has a faint extended, linear feature along the projected minor axis of its nuclear disk; 
this is the third structural component for the nucleus. 
We attribute it to the previously inferred bipolar outflow.

\newpage
\acknowledgements
We are grateful to the ALMA Observatory and its data archive team for the observations and data we used here.
We thank the ALMA help desk for their prompt response to our numerous inquiries about ALMA observations and data reduction.
We also thank our referee for comments that helped clarify this paper.
This paper makes use of the following ALMA data: ADS/JAO.ALMA\#2015.1.00113.S and  ADS/JAO.ALMA\#2011.0.00001.CAL.
ALMA is a partnership of ESO (representing its member states), NSF (USA) and NINS (Japan), 
together with NRC (Canada), MOST and ASIAA (Taiwan), and KASI (Republic of Korea), 
in cooperation with the Republic of Chile. 
The Joint ALMA Observatory is operated by ESO, AUI/NRAO and NAOJ.
This research has made use of NASA's Astrophysics Data System Bibliographic Services.
This research has also made use of the NASA/IPAC Extragalactic Database (NED), 
which is operated by the Jet Propulsion Laboratory, California Institute of Technology, 
under contract with the National Aeronautics and Space Administration.
KS is supported by MOST grants 105-2119-M-001-036 and 106-2119-M-001-025.

\vspace{0.5cm}
\facility{ALMA}
\software{CASA v4.7.2 \citep{CASA07}, mpfit \citep{More77,More93,Markwardt09}}

\bigskip

\end{document}